\documentclass[a4paper,fleqn]{cas-dc}

\usepackage[numbers]{natbib}

\def\tsc#1{\csdef{#1}{\textsc{\lowercase{#1}}\xspace}}
\tsc{WGM}
\tsc{QE}

\begin{document}
\let\WriteBookmarks\relax
\def\floatpagepagefraction{1}
\def\textpagefraction{.001}

\shorttitle{}    

\shortauthors{}  

\title [mode = title]{Malware Classification Based on Image Segmentation}  

\tnotemark[1] 

\tnotetext[1]{The source code and data covered in this article are available at https://github.com/Wanhu-Nie/Malware-Classification-Based-on-Image-Segmentation.} 

%

\author[1]{Wanhu Nie}[orcid=0009-0007-7304-8813]

\cormark[1]


\ead{wanhu.nie@qq.com;232081201004@lut.edu.cn}



\affiliation[1]{organization={Lanzhou University of Technology},
            city={Lanzhou},
            postcode={730000}, 
            state={Gansu},
            country={China}}





\credit{Conceptualization, Methodology, Software, Validation, Formal analysis, Investigation, Data Curation, Writing - Original Draft, Visualization}


\cortext[1]{Corresponding author}



\begin{abstract}
Executable programs are highly structured files that can be recognized by operating systems and loaded into memory, analyzed for their dependencies, allocated resources, and ultimately executed. Each section of an executable program possesses distinct file and semantic boundaries, resembling puzzle pieces with varying shapes, textures, and sizes. These individualistic sections, when combined in diverse manners, constitute a complete executable program. This paper proposes a novel approach for the visualization and classification of malware. Specifically, we segment the grayscale images generated from malware binary files based on the section categories, resulting in multiple sub-images of different classes. These sub-images are then treated as multi-channel images and input into a deep convolutional neural network for malware classification. Experimental results demonstrate that images of different malware section classes exhibit favorable classification characteristics. Additionally, we discuss how the width alignment of malware grayscale images can influence the performance of the model.
\end{abstract}





\maketitle

\section{Introduction}
\label{sec1}

Economic interests are a significant driving force behind the creation of malware. Criminals produce malicious software to steal users' personal information, sell it, or carry out online fraud. Alternatively, malware displays advertisements through popups, forced web page visits, or other means to generate economic benefits. Due to the increased ease of malware creation and distribution, the number of malware instances has exploded. According to a security report released by Rising Antivirus \cite{Rising}, the company's security platform intercepted 655,900 samples of ransomware in 2023, representing a 13.24\% increase compared to 2022.

As the threat of malware proliferates, security software has emerged as a response. Security software provides a comprehensive security framework capable of detecting and preventing malware invasions in real-time. However, money remains a powerful temptation, and the battle between malware defense and attack has become a protracted war. To evade detection, malware creators have developed various anti-detection techniques, such as encryption and obfuscation. They accomplish this by either adding encrypted shells to malware or encrypting the payload using encryption algorithms. The obfuscation methods employed by malware authors are also diverse, including the insertion of redundant code and alteration of instruction sequences \cite{You_Yim_2010,Alsmadi_Alqudah_2021,Sung_Xu_Chavez_Mukkamala_2005}.

To address this increasingly serious challenge, effectively classifying malware and accurately identifying the characteristics of its families is crucial. Malware analysis is divided into static analysis and dynamic analysis. Static analysis is the process of analyzing executable programs without running the code, which often requires reverse compiling the executable file. It determines behavioral intentions by analyzing the assembly instruction sequences of the executable program, thereby identifying whether it contains malicious code. While static analysis avoids the unknown risks associated with executing malware, it also faces challenges. For instance, static analysis is susceptible to the effects of malware encryption techniques and code obfuscation \cite{Moser_Kruegel_Kirda_2007}. Dynamic analysis of malware, on the other hand, requires running the code, necessitating the use of sandbox or virtual machine environments to simulate a real-world system environment \cite{Jamalpur_Navya_Raja_Tagore_Rao_2018}. Dynamic analysis discovers malware by monitoring behavioral characteristics such as network communication, file read/write operations, and others \cite{Rieck_Holz_Willems_Düssel_Laskov_2008,Pirscoveanu_Hansen_Larsen_Stevanovic_Pedersen_Czech_2015,Boukhtouta_Mokhov_Lakhdari_Debbabi_Paquet_2016}. However, this detection method requires an additional virtual execution environment, resulting in higher costs \cite{Bayer_Kirda_Kruegel_2010}. Furthermore, some malware has the ability to detect virtual environments. Once malware detects that it is operating in a virtual environment, its behavior may change or terminate execution, adding complexity to the malware analysis process.

In the face of the ever-increasing number of malware, traditional malware detection methods are struggling to keep up. In recent years, deep learning has achieved remarkable success in fields such as computer vision and natural language processing. Deep learning has also been employed to tackle the issue of malware classification \cite{Nataraj,Kalash_Rochan_Mohammed_Bruce_Wang_Iqbal_2018,Vasan_Alazab_Wassan_Naeem_Safaei_Zheng_2020}. In the field of malware visualization classification, the binary sequences of malware are transformed into pixel arrays of grayscale images, thus converting the malware classification problem into an image classification problem \cite{Bijitha_Nath_2022}. Compared to traditional malware classification methods that rely on domain knowledge, the use of deep learning for malware visualization classification significantly reduces the workload of feature engineering.

\begin{figure}
	\centering
	\includegraphics[width=1\columnwidth]{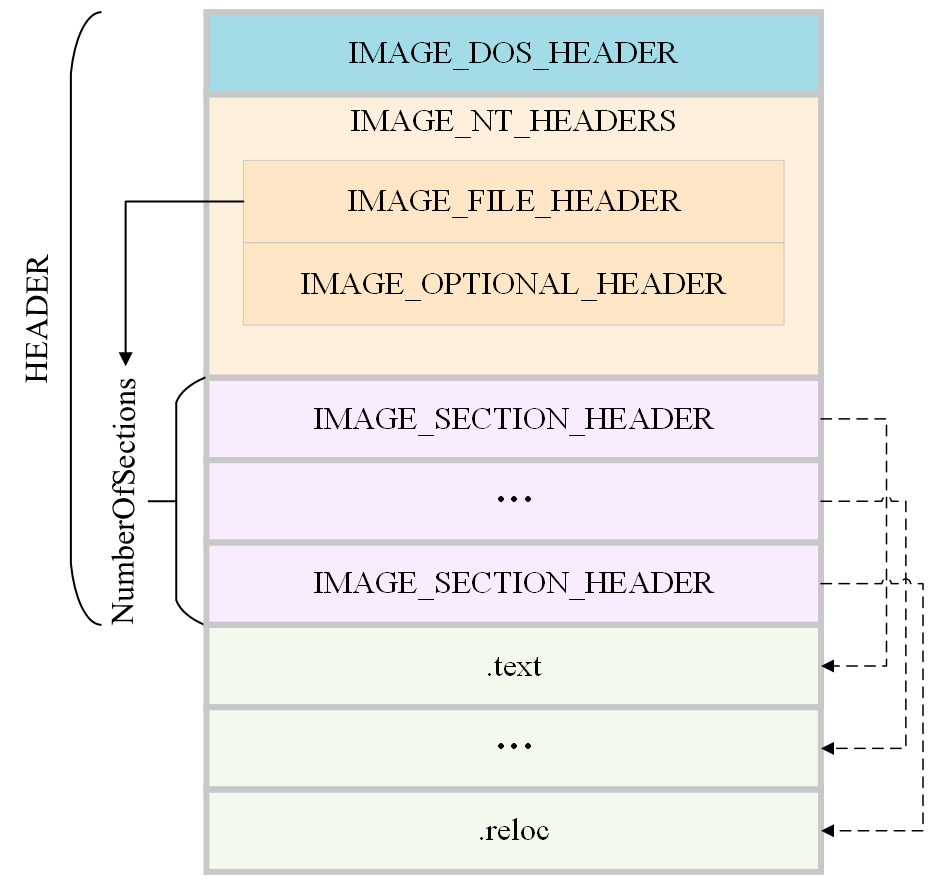}
    \caption{Correspondence between section table and sections \cite{PE-format}. In the PE-Header of Windows executable programs, the section table describes the spatial layout of the entire executable, including information such as the number, order, and size of sections. Each section carries specific types of code or data with independent attributes. Executable programs exhibit distinct file and semantic boundaries.}
    \label{fig1}
\end{figure}

In this paper, considering the natural file boundaries and semantic boundaries of executable programs (as shown in \autoref{fig1}), we propose a novel and challenging approach for malware visualization classification. Specifically, we attempt to segment the grayscale images of malware based on the section categories, generating multiple sub-images of different types. These sub-images are then treated as multi-channel images and fed into a deep convolutional neural network for classification. This approach offers a fresh perspective on malware visualization classification. Furthermore, we discuss how the width of the images can impact the performance of malware classification models based on image visualization.

\section{Related work}
Nataraj et al. \cite{Nataraj} pioneeringly proposed converting the binary sequences of malware into grayscale images for malware classification, avoiding complex operations such as decompilation and analysis of the malware. Kalash and Rochan et al. \cite{Kalash_Rochan_Mohammed_Bruce_Wang_Iqbal_2018} proposed using a fine-tuned CNN (Convolutional Neural Network) to classify visualized malware and achieved good results. Vasan et al. \cite{Vasan_Alazab_Wassan_Naeem_Safaei_Zheng_2020} proposed converting grayscale images of malware into color images and achieved malware classification through a fine-tuned CNN network. However, malware authors can obfuscate their creations by altering the order of sections or inserting large amounts of redundant data within sections, which can change the partial texture and spatial layout features of the malware grayscale images. These classification methods based on malware visualization may face significant challenges or even be defeated \cite{Nataraj}.

The number, order, and size of sections are crucial features in executable programs. Due to the file alignment mechanism, the length of each section is required to be aligned according to the FileAlignment field in the IMAGE\_OPTIONAL\_HEADER, resulting in the tail of each section being padded with several zeros. Since the number of paddings at the end of each section is not fixed, the boundaries between some sections may not be clear enough. Xiao et al. \cite{Xiao_Guo_Shen_Cui_Jiang_2021} proposed adding a colored label box for each section to enhance the distribution information of the sections, where the width of the border is determined by the length of the section. However, introducing label boxes will cover some pixels of the section. Additionally, Chaganti et al. \cite{Chaganti_Ravi_Pham} found that using grayscale images of malware with a fixed width is beneficial for malware classification, but they did not analyze the reasons for the effectiveness of this strategy.

Compared to related work, our proposed method effectively overcomes the potential issue of ambiguous boundaries between sections in executable programs, allowing the model to learn the texture characteristics of different types of sections in executable programs. Furthermore, this section-based segmentation strategy exhibits spatial independence, focusing on the distribution patterns of malware sections rather than the overall original image structure. This is because the spatial layout of malware may be reconstructed by attackers, potentially misleading the classifier.

\section{Segmentation of malware grayscale images}
\begin{figure*}
	\centering
	\includegraphics[width=1\textwidth]{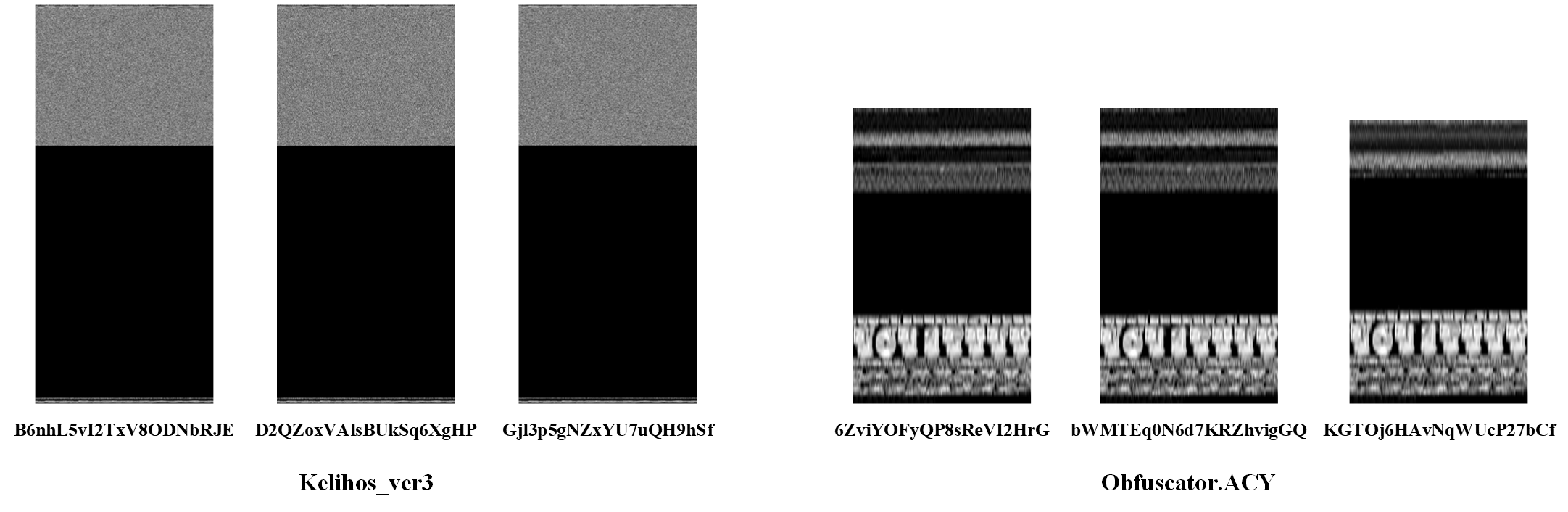}
    \caption{Grayscale image comparison of malware belonging to the same family. These malware samples are from the Microsoft Malware Dataset \cite{malware-classification}, which only shows samples from two malware families for comparison.}
    \label{fig2}
\end{figure*}

Malware grayscale images from the same family tend to have similar spatial layouts and texture characteristics \cite{Nataraj}, while those from different families are well distinguished, which serves as the basis for image-based malware classification methods, as shown in \autoref{fig2}.

Malware and its variants from the same family tend to utilize similar codes and data \cite{Nataraj_Karthikeyan_Manjunath_2015}, which manifest as similar textures in the visualized images. During the program linking phase, if no specific linking script is provided, the linker will follow its internal algorithm to link the program, resulting in similar spatial layouts among the linked programs. However, to evade detection by anti-malware solutions, malware authors often employ techniques such as obfuscation or encryption to disguise their malware as benign software. For instance, they may alter the order of sections, or insert significant amounts of redundant data into sections. These manipulations can alter the texture and even the spatial layout of the malware grayscale images, leading to more complex and diverse image representations. If the model can learn the texture features and spatial layout patterns of various section types, while demonstrating greater robustness to alterations in malware's structure and texture features, this would be of significant importance for the task of image-based malware classification.

\begin{figure*}
	\centering
	\includegraphics[width=1\textwidth]{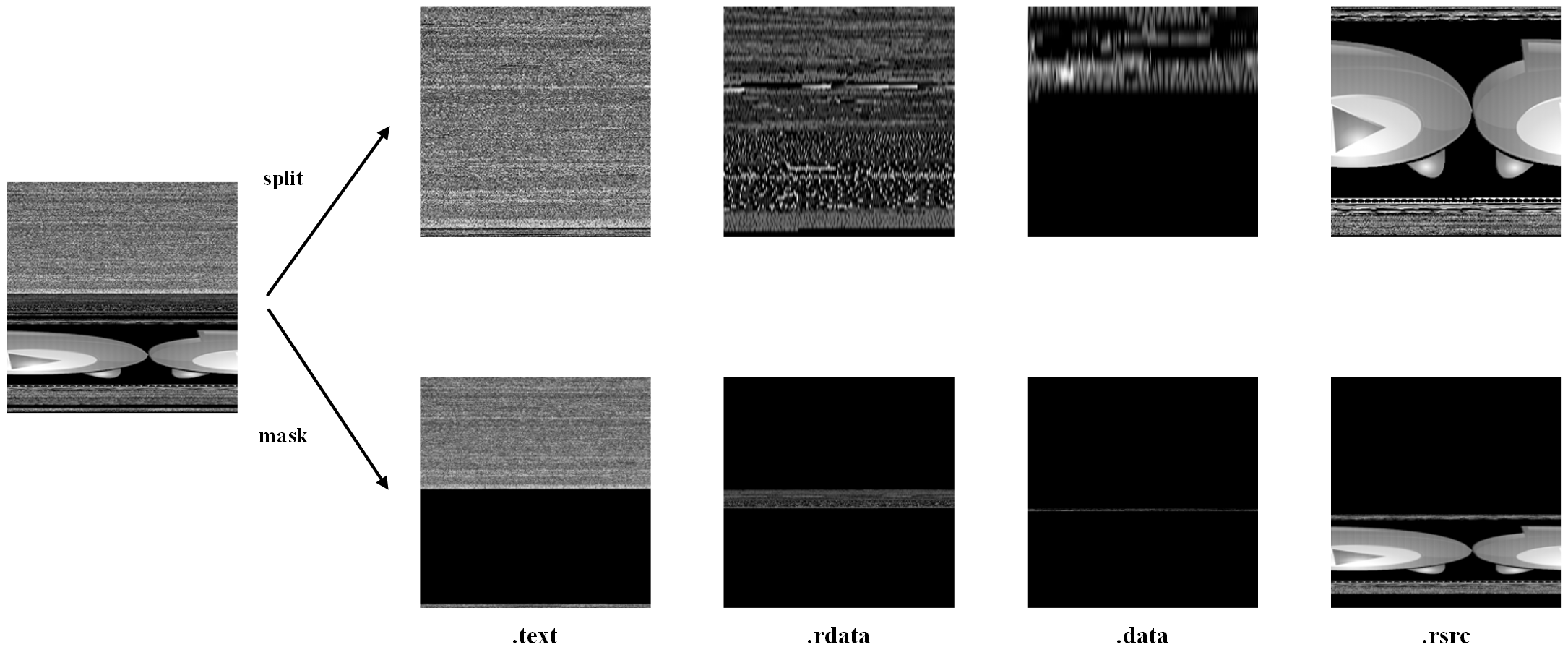}
    \caption{Two approaches for segmentation of malware grayscale images.}
    \label{fig3}
\end{figure*}

The proposed method of segmenting malware grayscale images based on section categories in this paper is subdivided into two approaches, as illustrated in \autoref{fig3}.

\begin{itemize}
\item Split: This approach segments the grayscale image based on section categories, resulting in multiple sub-images. This method ignores the spatial layout of the malware, allowing the model to focus on learning the texture features of the sub-image corresponding to each section.

\item Mask: After segmenting the grayscale image based on section categories, this approach only displays the pixels corresponding to the respective section in each sub-image, while pixels belonging to other sections are masked out. This method preserves the spatial layout information of the sections, enabling the model to consider the distribution characteristics of the sections while learning the image texture features.
\end{itemize}

When segmenting the grayscale images of malware, it's important to note that malware may obfuscate section names or attributes, such as disguising the .text section to a non-standard name or setting its attributes to make it appear as a read-only code segment. Therefore, the type of a section cannot be solely determined by its name. Instead, a combination of the section's name and attributes should be used to decide its type. \autoref{tab1} provides a list of section names and their corresponding attributes.

\begin{table*}[width=0.57\textwidth]
  \centering
  \caption{Section and its attributes \cite{PE-format}.}
  \label{tab1}
  \begin{tabular}{lll}
    
    \toprule
    Section & Content & Characteristics	\\
    \midrule
	.text & Executable code & Code, Read/Execute \\
	.rdata & Read-only initialized data & Initialized data, Read \\
	.data & Initialized data & Initialized data, Read/Write \\
	.edata & Export tables & Initialized data, Read \\
	.idata & Import tables & Initialized data, Read/Write \\
	.tls & Thread-local storage & Initialized data, Read/Write \\
	.bss & Uninitialized data & Uninitialized data, Read/Write \\
	.reloc & Image relocations & Initialized data, Read \\
	.rsrc & Resource directory & Initialized data, Read \\
    \bottomrule
  \end{tabular}
\end{table*}

\section{Malware classification}
\subsection{Dataset}
The Microsoft Malware Dataset (BIG 2015) \cite{malware-classification,Ronen_Radu_Feuerstein_Yom-Tov_Ahmadi_2018} was released by Microsoft in 2015 during a malware classification competition hosted on Kaggle. The dataset is divided into a training set and a test set, containing 10,868 and 10,873 malware samples respectively, belonging to 9 different malware families. Each sample consists of two files: a .bytes file containing the binary content of the malware sample (excluding the PE-Header), and a .asm file generated by IDA, which is a disassembly of the binary file. It's worth noting that the samples in the test set do not have labels.

\begin{table}
  \caption{Sample distribution of Microsoft Malware Dataset (training set).}
  \label{tab2}
  \centering
  \begin{tabular}{lr}
    \toprule
    Family & Samples \\
    \midrule
	Ramnit & 1541 \\ 
	Lollipop & 2478 \\
	Kelihos\_ver3 & 2942 \\
	Vundo & 475 \\
	Simda & 42 \\
	Tracur & 751 \\
	Kelihos\_ver1 & 398 \\
	Obfuscator.ACY & 1228 \\
	Gatak & 1013 \\
	\midrule
	Total & \textbf{10868} \\
    \bottomrule
  \end{tabular}
\end{table}

Since the .bytes files of malware samples do not contain the PE-Header, the section table cannot be extracted directly from them. However, the .asm files, which are generated by disassembling the binary files using IDA, contain markers at the beginning of each section, indicating crucial features such as the section's attributes and length. Therefore, section information can be extracted from the .asm files. In this experiment, we use the .bytes files combined with the section information to generate the complete grayscale images of the malware as well as the segmented sub-images based on the section types.

\subsection{Data Processing}
\begin{figure}
	\centering
	\includegraphics[width=1\columnwidth]{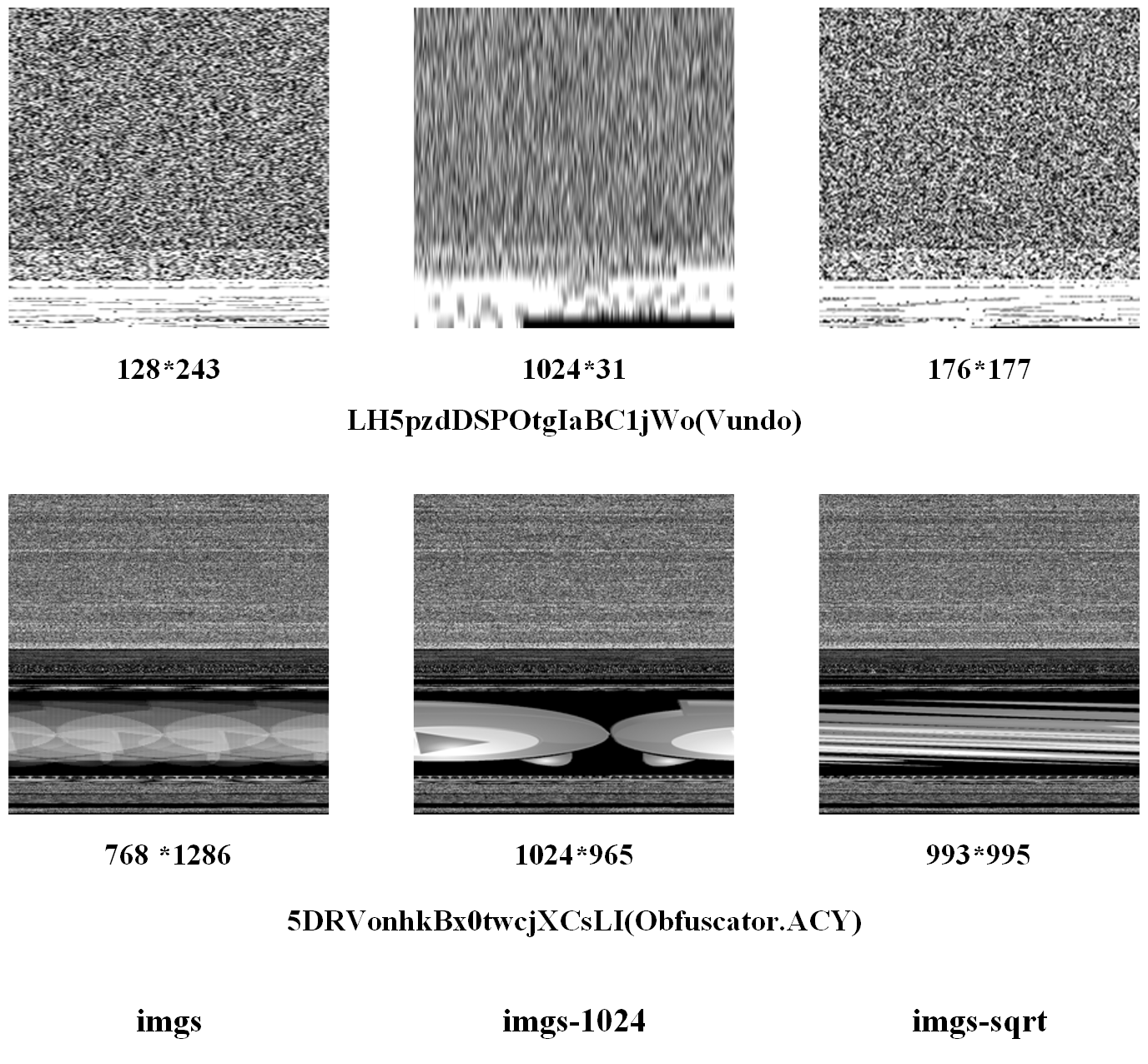}
    \caption{Impact of width alignment on grayscale images of malware. The grayscale images of malware shown are the results after scaling.}
    \label{fig4}
\end{figure}

Malware samples can vary significantly in size, ranging from just a few dozen kilobytes for smaller ones to over 100 megabytes for larger ones. When employing image-based classification methods, it's crucial to resize the grayscale images of the malware to a uniform size. The choice of image width not only affects the degree of detail loss but also alters the texture and structural features of the image. For larger malware, resizing their grayscale images can result in the loss of local details, preserving only the general texture and structural characteristics. On the other hand, for smaller malware, the grayscale images may become distorted due to stretching, and the alignment method can have a more significant impact on the texture features, as shown in \autoref{fig4}. Notably, when the .rsrc section contains resources like images, the alignment of the image width can have a particularly pronounced effect on the grayscale image, potentially causing significant differences even among images with originally similar textures. Therefore, the alignment method for image width emerges as a new hyperparameter in the problem of malware visualization and classification \cite{Raff_Barker_Sylvester_Brandon_Catanzaro_Nicholas_2018}.

Nataraj et al. \cite{Nataraj} proposed a scheme in their paper for aligning the width of malware grayscale images based on their file size, which has been widely adopted in subsequent research on malware visualization and classification. Chaganti et al. demonstrated that using a fixed width (512 bytes) for malware grayscale images is beneficial for classification tasks, at least in the context of the Microsoft Malware Dataset \cite{malware-classification}.

\begin{table}
  \caption{Width alignment scheme for malware grayscale images proposed by Nataraj et al.}
  \label{tab3}
  \centering
  \begin{tabular}{ll}
    \toprule
    File size range & Image width \\
    \midrule
	<10KB & 32 \\
	10KB-30KB & 64 \\
	30KB-60KB & 128 \\
	60KB-100KB & 256 \\
	100KB-200KB & 384 \\
	200KB-500KB & 512 \\
	500KB-1000KB & 768 \\
	>1000KB & 1024 \\
    \bottomrule
  \end{tabular}
\end{table}

In this paper, we utilize the Microsoft Malware Dataset to construct five different image-based malware datasets for comparative experiments. Our goal is to validate the effectiveness of the proposed method and investigate how different alignment schemes for the grayscale image width can influence the model's performance.

\begin{itemize}
\item {S1: Using the scheme described in \autoref{tab3}, where the width of the grayscale image of the malware is aligned based on the range of its file size.}
\item {S2: Aligning the width of the malware grayscale image based on the square root of its file size, which introduces variety in the width alignment of the dataset samples.}
\item {S3: Uniformly aligning the width of the malware grayscale image to 1024 bytes.}
\item {S4: Segmenting the grayscale image based on section types, resulting in multiple sub-images.}
\item {S5: Segmenting the grayscale image based on section types, where each sub-image displays only the pixels corresponding to the respective section, while the pixels of other sections are masked out.}
\end{itemize}  

In the .bytes files, bytes represented as “??” are uniformly converted to 0 instead of being discarded directly. This strategy is applied in all five schemes, S1 to S5. For S4 and S5, the grayscale images are uniformly aligned to a width of 1024 bytes.

In S4 and S5, the section types are categorized based on HEADER, .text, .rdata, .data, .rsrc, and others. Sections such as .bss, .tls, and .idata are grouped into “others” because they are uninitialized in PE files or have limited lengths, making it difficult to extract meaningful texture features. It's worth noting that, due to security considerations, the samples in the Microsoft Malware Dataset do not contain the PE-Header. Therefore, for the samples in this dataset, the “HEADER” category is effectively invalid.

\subsection{Model overview}
In this work, we utilize existing deep convolutional neural networks to classify the visualized malware, specifically VGG16 and ResNet50. VGG16 \cite{Simonyan_Zisserman_2015}, a deep convolutional neural network, demonstrates exceptional performance in large-scale image recognition tasks through the stacking of multiple convolutional layers, paving the way for subsequent deep learning models to develop deeper network architectures. VGG16 employs 3x3 convolutional kernels to construct the network, enhancing the model's nonlinear transformation capabilities and improving its generalization ability. Additionally, Residual Networks (ResNet) \cite{He_Zhang_Ren_Sun_2016} introduce residual connections, which enable the model to preserve some of the input data's information while extracting features. This design allows gradients to propagate to deeper layers of the network, effectively addressing the issue of gradient vanishing and enabling the network to learn the data's features better. Among the ResNet family, ResNet50 strikes a good balance between model performance and computational resource requirements, compared to other models like ResNet18 and ResNet101.

\begin{figure}
	\centering
	\includegraphics[width=0.5\columnwidth]{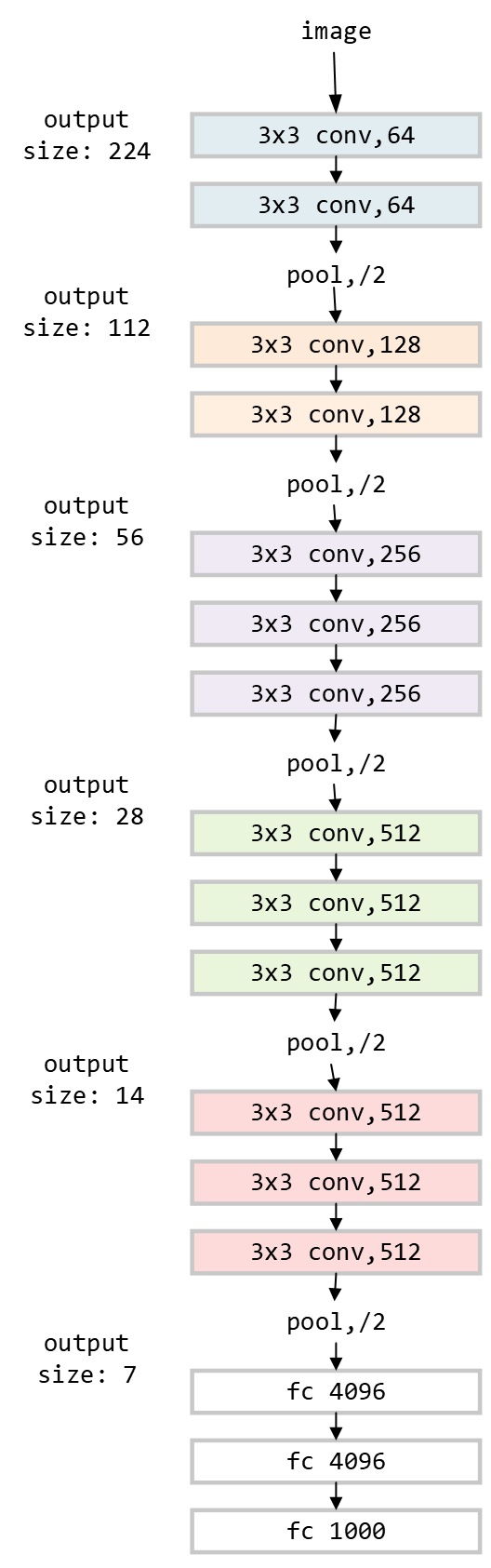}
    \caption{Architecture of VGG16.}
    \label{fig5}
\end{figure}

\begin{figure*}
	\centering
	\includegraphics[width=0.16\textwidth,angle=90]{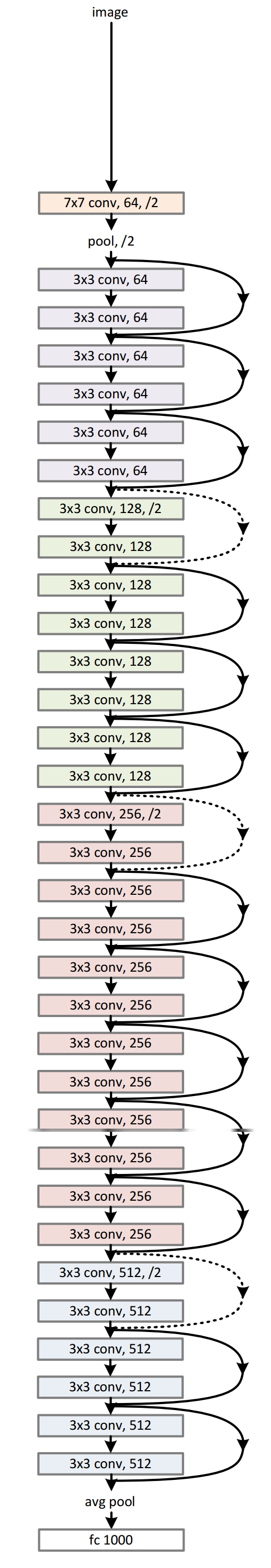}
    \caption{Architecture of ResNet50 \cite{He_Zhang_Ren_Sun_2016}.}
    \label{fig6}
\end{figure*}

We directly utilize pre-trained models provided by Torchvision for transfer learning. To accommodate the scenario where the number of input image channels exceeds 3, the two-dimensional convolutional input channels of the input layer in the deep convolutional neural network are modified accordingly. We conducted experiments using the VGG16 and ResNet50 models on the datasets to evaluate the effectiveness of the proposed method in this paper.

\subsection{Experiments and evaluationsn}
All experiments were conducted on a 64-bit Ubuntu 20.04.5 LTS system, utilizing a Xeon(R) Platinum 8255C CPU (2.50GHz) and an NVIDIA RTX 2080 Ti GPU (11GB). Due to the limitation of GPU memory, the batch size for ResNet was set to 64. We employed the cross-entropy loss function to train the models and used stochastic gradient descent (SGD) to optimize the model parameters. ResNet50 utilized an exponential decay learning rate scheduling strategy with a decay rate of 0.9 to reduce fluctuations during the later stages of training and enhance model performance. To maximize the model's generalization ability and prevent overfitting, the order of samples was shuffled at the end of each epoch during training. The parameter settings of the models are detailed in \autoref{tab4}.

\begin{table*}[width=0.98\textwidth]
  \caption{Model parameter settings.}
  \label{tab4}
  \centering
  \begin{tabular}{lllllll}
    \toprule
    Model & Epochs & Batch size & Initial learning rate & Learning rate scheduling strategy & Weight decay & Momentum \\
    \midrule
	VGG16 & 20 & 8 & 0.001 & false & 0.0005 & 0.9 \\
	ResNet50 & 15 & 64 & 0.01 & true (0.9) & 0.006 & 0.9 \\
    \bottomrule
  \end{tabular}
\end{table*}

In this work, we use accuracy (\%) to evaluate the model's fit on the training set, representing the percentage of malware correctly classified by the model. Deep convolutional neural networks like ResNet50, due to their powerful representation capabilities, can easily achieve very high accuracy on the training set, even reaching 100\%. However, this does not necessarily indicate good generalization ability and may suggest overfitting. Therefore, it is crucial to evaluate the model's performance on a test set in the context of malware visualization classification. The predictions on the test set are submitted to an online evaluation server \cite{malware-classification}, which assesses the predictions using multi-class logarithmic loss and returns a logloss score comprising both Private Score and Public Score. Here, the Private Score is used as the final result, calculated using 70\% of the test set samples, while the Public Score is computed using the remaining 30\% of the samples. The multi-class logarithmic loss function is defined as follows:
\[logloss = - \frac{1}{N} \sum_{i=1}^{N} \sum_{j=1}^{M} y_{ij} \log(p_{ij})\]
Where \(N\) represents the number of samples in the test set, \(M\) represents the number of malware families, and log denotes the natural logarithm. If sample \(i\) belongs to malware family \(j\), then \(y_{ij}\) is 1, otherwise, it is 0. \(p_{ij}\) represents the predicted probability that sample \(i\) belongs to malware family \(j\).

By evaluating the model's performance on the test set using this multi-class logarithmic loss, we can gain a more accurate understanding of the model's generalization ability and how well it can classify unseen malware samples. The lower the logloss score, the better the model is at predicting the correct malware family for each sample.

\begin{table*}[width=0.82\textwidth]
  \caption{Experimental results on Microsoft Malware Dataset.}
  \label{tab5}
  \centering
  \begin{tabular}{lllllll}
    \toprule
    Model & Dataset(,tag) & Accuracy (training set) & Logloss (test set) \\
    \midrule
	\multirow{6}{*}{VGG16} & S1 & 99.98\% & 0.0401  \\
	& S2 & 99.94\% & 0.0547  \\
	& S3 & 99.94\% & 0.0411  \\
	& S4(.text+.rdata+.rsrc) & 99.95\% & 0.0577  \\
	& S5(.text+.rdata+.rsrc) & 99.84\% & 0.0889  \\
	& S5(imgs-1024+.text+.rsrc) & 99.99\% & 0.0521  \\
	\midrule
	\multirow{14}{*}{ResNet50} & S1 & 100\% & 0.0316  \\
	& S2 & 100\% & 0.0330  \\
	& S3 & 100\% & \textbf{0.0265}  \\
	& S4(.text+.data+.rsrc) & 99.99\% & 0.0383  \\
	& S4(.text+.rdata+.rsrc) & 99.97\% & 0.0320  \\
	& S4(.text+.rdata+.data) & 99.93\% & 0.0452  \\
	& S4(.rdata+.data+.rsrc),T1 & 99.30\% & 0.0587  \\
	& S4(.text+.rdata+.data+.rsrc) & 99.91\% & 0.0364  \\
	& S5(.text+.rdata+.rsrc) & 99.95\% & 0.0446  \\
	& S5(.text+.rdata+.data+.rsrc) & 99.93\% & 0.0505  \\
	& S5(imgs-1024+.text+.rsrc),T2 & 99.98\% & 0.0279  \\
	& S5(imgs-1024+.text+.rdata) & 100\% & 0.0292  \\
	& S5(imgs-1024+.text+.data) & 99.98\% & 0.0286  \\
	& S5(imgs-1024+.text+.rdata+.data+.rsrc) & 99.97\% & 0.0385 \\
    \bottomrule
  \end{tabular}
  \\ \raggedright \small (S1 to S3 expand the single channel of grayscale images to 3 channels, while S4 and S5 treat multiple sub-images as a multi-channel image.)
\end{table*}

\autoref{tab5} presents the performance of the models on the Microsoft Malware Dataset. For the S2 dataset, due to the diverse alignment of sample widths, its performance on both VGG16 and ResNet50 is inferior to S1 and S3. Similarly, S3 exhibits significantly better performance on ResNet50 compared to S1 and S2, achieving a logloss of 0.0265 on the test set. This indicates that, at least in the Microsoft Malware Dataset, the alignment method of grayscale image widths can impact the model's performance. Employing a fixed width alignment strategy helps mitigate the impact of factors such as distortion, texture changes, and structural feature alterations that may occur during image scaling.

On ResNet50, S4 exhibits similar performance to S1-S3. This suggests that in the visualization-based classification of malware, even when ignoring the spatial layout of the executable, segmenting grayscale images based on section types can still yield good results. Since the sub-images in S5 are generated by masking the original image according to section types, they inevitably face issues of sparse image pixels, leading to poor model performance. However, when combined with the single-channel original image, S5 demonstrates good performance. For example, the 3-channel image combination method in T2 achieves a logloss of 0.0279.

The comparative experiments between S4 and S5 intuitively demonstrate that different types of sections contribute differently to the classification task. The .text section, being the most crucial part of an executable, has a significant impact on the model's performance whether it is included in the classification. However, even in the absence of the .text section, the model can still show satisfactory performance, such as in T1. This indicates that malware from the same family tends to have similar data structures and string patterns. Furthermore, the .rsrc section, which stores the program's resource files, exhibits excellent classification characteristics. This means that malware from the same family and its variants tend to use similar resource files, such as icons, menus, and dialog boxes.

\section{Limitations}
Although the spatial layout of malware can be restructured by attackers, it still holds valuable information. However, to avoid the spatial dependencies of the entire grayscale image of malware (rather than between sections), S4 only extracts sub-images of each section, overlooking the spatial layout features. In contrast, S5 shows only the pixel points of a specific section type while masking the rest, preserving part of the spatial layout features but also introducing sparsity into the images.

Furthermore, since S4 and S5 extract sub-images from multiple sections and treat them as multi-channel images for training, it poses numerous challenges for the migration of pre-trained models. As the images in different channels belong to different types of sections, significant differences exist between the channels, which can affect the model's ability to extract effective features. Pre-trained models are designed specifically for 3-channel images, and attempting to change the number of input channels disrupts the model's dependencies, potentially limiting its ability to learn meaningful features. Additionally, increasing the number of channels introduces more noise and increases model complexity. This requires more training techniques and optimized model architecture to further improve performance.

\section{Conclusions}
This paper proposes a malware classification method based on image segmentation according to section types. Initial experimental results indicate that this approach can achieve performance comparable to traditional image-based malware classification methods. Due to its spatial independence, this method holds promise for demonstrating greater robustness against obfuscation techniques that alter malware textures and spatial layouts, but further validation is needed in future work. Additionally, this method allows the model to learn texture features from different types of sections in executable programs, opening up more possibilities for future research in the field of image-based malware classification.

Inspired by the attention mechanism, it may be possible to generate a mask matrix based on the original malware grayscale image and its section information to describe the types of different sections in the grayscale image. Then, by applying the attention mechanism, we can emphasize important regions while suppressing less significant ones. This is a potentially valuable approach, but further experiments are needed to validate its effectiveness.

In the broader context of image-based malware classification, the width of malware grayscale images emerges as a novel hyperparameter that directly impacts the model's performance. Due to the variance in malware sizes, alterations in image texture and structure, as well as the loss of detail and distortion that occurs during image scaling, these issues are closely tied to the alignment method of the image width. The root of the problem lies in the fact that binary files are one-dimensional sequences, rather than true two-dimensional images. Therefore, exploring ways to optimize the alignment of malware grayscale image widths will be an important area of future work.









\printcredits

\section*{Acknowledgments}
This research did not receive any specific grant from funding agencies in the public, commercial, or not-for-profit sectors.

\bibliographystyle{cas-model2-names}

\bibliography{cas-refs}



\end{document}